\documentclass[aps,showpacs,twocolumn]{revtex4}
\usepackage[dvips]{graphics,graphicx}
\usepackage{amsmath}
\usepackage{amssymb}                 
\usepackage{amscd}                   
\usepackage{wasysym}
\newcommand{\ri}{{ \rm i }}

\newcommand{\rd}{{ \rm d }}

\usepackage[ansinew]{inputenc}
\usepackage[T1]{fontenc}
\usepackage{ae,aecompl}

\newcommand{\be}{\begin{equation}}
\newcommand{\ee}{\end{equation}}

\newcommand{\nn}{\nonumber}

\DeclareGraphicsExtensions{.eps}
\begin{document}
\bibliographystyle{apsrev}
\title{Mean-field dynamics of a Bose-Einstein condensate in a time-dependent triple-well trap:
Nonlinear eigenstates, Landau-Zener models and STIRAP}
\author{E. M. Graefe, H. J. Korsch, and D. Witthaut}
\email{witthaut@physik.uni-kl.de}
\affiliation{Technische Universit{\"a}t Kaiserslautern, FB Physik,
                            D-67653 Kaiserslautern, Germany}
\date{\today }

\begin{abstract}
We investigate the dynamics of a Bose--Einstein condensate (BEC)
in a triple-well trap in a three-level approximation.
The inter-atomic interactions are taken into account in
a mean-field approximation (Gross-Pitaevskii equation),
leading to a nonlinear three-level model.
New eigenstates emerge due to the nonlinearity, depending
on the system parameters. Adiabaticity breaks down if such
a nonlinear eigenstate disappears when the parameters are varied.
The dynamical implications of this loss of adiabaticity are
analyzed for two important special cases:
A three level Landau-Zener model and the STIRAP scheme.
We discuss the emergence of looped levels for an equal-slope
Landau-Zener model. The Zener tunneling probability does not
tend to zero in the adiabatic limit and shows pronounced oscillations
as a function of the velocity of the parameter variation.
Furthermore we generalize the STIRAP scheme for adiabatic coherent
population transfer between atomic states to the nonlinear case.
It is shown that STIRAP breaks down if the nonlinearity exceeds
the detuning.
\end{abstract}

\pacs{03.75.Lm, 03.65.Kk, 32.80.Qk,  42.65.Sf \\
Keywords: Bose-Einstein condensates, Gross-Pitaevskii, Zener tunneling, STIRAP, coherent control}
\maketitle


\section{Introduction}
\label{sec-intro}

The experimental progress in controlling Bose-Einstein condensates
(BECs) has led to a variety of spectacular results in the last few years.
At very low temperatures, the dynamics of a BEC can be described in a
mean-field approximation by the Gross-Pitaevskii (GPE) or nonlinear
Schr\"odinger equation (NLSE) \cite{Pita03}.
Previously, several authors investigated the dynamics of the NLSE for a
double-well potential in a two-mode approximation
\cite{Smer97,Milb97,Zoba00,Wu00,Holt01b,Dago02,Wu03,05catastrophe}.
Novel features were found, e.g.~the emergence of new nonlinear stationary
states \cite{Dago02} and a variety of new crossing scenarios
(cf.~\cite{05catastrophe} and references therein).
Studies of the quantum dynamics beyond mean-field theory were reported
in \cite{Smer97,Holt01b,Vard01}.
First approaches of the coherent control of BECs in driven double-well
potentials have been reported \cite{Holt01b}.
Another relevant application of the two-level NLSE is the
dynamics of a BEC in an accelerated or tilted optical lattice \cite{Wu00,Wu03}.
Furthermore, the NLSE describes the propagation of light pulses in nonlinear
media \cite{Dodd82} and its two-level analogon has been studied in the context of
a nonlinear optical directional coupler \cite{Chef96}.
This equation is also known as the discrete self-trapping equation and has
been applied, e.g., to the dynamics of quantum dimers. In fact, the
characteristic loop structures \cite{Wu00} important in the following,
have been already observed in this context \cite{Esse95}.

In the present paper we will extend these studies to nonlinear three-level
quantum systems with special respect to the breakdown of adiabatic evolution
due to the nonlinearity.
In fact, we investigate the dynamics of a BEC in a three-level system
\be
  \psi(t) = a(t) \psi_1 + b(t) \psi_2 + c(t) \psi_3
\ee
in a mean-field approach.
The dynamics of the coefficients is described by the discrete NLSE
\be \label{dNLSE}
  H(|a|^2,|b|^2,|c|^2)
  \left(\begin{array}{c} a \\ b \\ c \end{array}\right)
   = \ri \hbar \frac{\rd}{\rd t} \left(\begin{array}{c} a \\ b \\ c \end{array}\right)
\ee
with the nonlinear three-level Hamiltonian
\be
  H(|a|^2,|b|^2,|c|^2) = \left(\begin{array}{c c c}
  \epsilon + g |a|^2 & v & 0 \\
  v & g |b|^2 & w \\
  0 & w & \delta + g |c|^2\\
  \end{array}\right) \, .
  \label{eqn-nlham-intro}
\ee
The dynamics preserves the normalization, which is fixed as
$|a|^2+|b|^2+|c|^2 = 1$. Throughout this paper units are rescaled
such that all properties become dimensionless and $\hbar = 1$.

\begin{figure}[t]
\centering
\includegraphics[width=8cm,  angle=0]{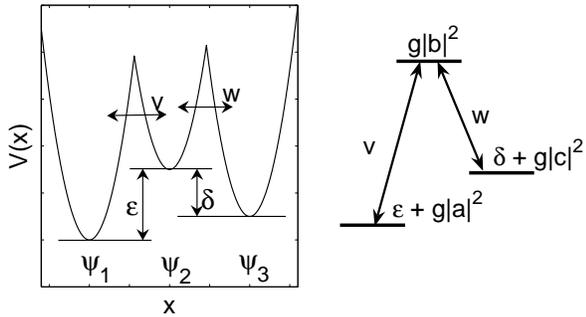}
\caption{\label{fig-triple-well}
Illustration of the triple-well potential and the corresponding
three-level model (\ref{eqn-nlham-intro}) investigated in the
present paper.}
\end{figure}

An experimental setup leading very naturally to the Hamiltonian
(\ref{eqn-nlham-intro}) is the dynamics of a BEC in a triple-well
potential. In this case the basis states $\psi_n(x)$ are localized
in the three wells. Then $\epsilon$ and $\delta$ are the on-site energies
of the outer wells and $v$ and $w$ denote the tunneling matrix elements
between the wells. These parameters can be varied by controlling the depth
or the separation of the wells. The energy scale is chosen such that
the on-site energy of the middle well is zero. This situation is
illustrated in figure \ref{fig-triple-well}.

A detailed discussion of these approximations and the validity of the
model can be found in \cite{Milb97} for the case of a double-well potential.
Furthermore, the existence of a dark state for ultracold atoms and molecules,
which was demonstrated experimentally very recently, could be
described successfully within a nonlinear three-level model
(see \cite{Wink05} and references therein)
Further applications include the dynamics of three-mode systems of
nonlinear optics or the analysis of the excitonic-vibronic coupled
quantum trimer.

As previously shown for the two-level system, the nonlinearity leads to the
emergence of new eigenstates without a linear counterpart, loop structures and
novel level crossing scenarios.
The concept of adiabaticity is very different in the nonlinear case
\cite{Liu03}, leading to nonlinear Zener tunneling \cite{Wu00} and
possibly to dynamical instability \cite{05catastrophe}.
These issues are discussed in section \ref{sec-zener} for the
three-level system (\ref{eqn-nlham-intro}).
Furthermore we analyze adiabatic coherent population transfer.
In the linear case a complete population transfer can be achieved using
the Stimulated Raman Adiabatic Passage (STIRAP) via a dark state of
the system \cite{Berg98b}.
Coherent control techniques for ultracold atoms in a triple-well
trap were previously investigated by Eckert et.~al.~for the case
of single atoms \cite{Ecke04}.
In the present paper it is shown that a mean-field interaction
plays a crucial role and that the STIRAP scheme fails if the nonlinearity
exceeds a critical value due to the breakdown of adiabaticity.

\section{Nonlinear Eigenstates}
\label{sec-eigenstates}

Nonlinear eigenstates and eigenvalues can be defined as the solutions
of the time-independent NLSE
\be
  H(|a|^2,|b|^2,|c|^2)
  \left(\begin{array}{c} a \\ b \\ c \end{array}\right)
   = \mu \left(\begin{array}{c} a \\ b \\ c \end{array}\right)
  \label{eqn-def-eigenstates}
\ee
with the chemical potential $\mu$.
Obviously an interpretation in terms of linear algebra is not feasible anymore.
Therefore -- although equation~(\ref{eqn-def-eigenstates}) looks like a common
eigenvalue-equation -- most of the implications of linear quantum mechanics are
not valid anymore. Nevertheless these eigenstates are of great importance for
understanding the dynamics of the system, as they still are stationary solutions
of the time dependent NLSE~(\ref{dNLSE}).

In the two-level case one can calculate the eigenstates by solving a
forth-order polynomial equation \cite{Wu00}. The case of three levels
turns out to be a bit more complicated.
As all parameters are real, the amplitudes $(a,b,c)$ are also real.
For almost all parameters (except in the ''uncoupling'' limits
$v,w \rightarrow 0$ or $\delta, \epsilon \rightarrow \infty$)
the amplitudes are all non-zero.
Thus the variables $x = b/a$ and $y = c/b$ are well defined
and determined by the equations
\begin{eqnarray}
&& (1-x^2 y^2) \left(g+\delta + \epsilon + v(x+x^{-1}) + w(y+y^{-1}) \right) \nn  \\
&& \quad - 3 x^2 +y^2 (\epsilon+vx)- 3 w y^{-1} + 3 \delta = 0
  \label{eqn-eigen-nl1}   \\
&& (1-x^2) \left(g+\delta + \epsilon + v(x+x^{-1}) + w(y+y^{-1}) \right) \nn \\
&& \quad - 3 x^{-1} - 3 w y + 3 x^2 (\epsilon +vx) = 0. \label{eqn-eigen-nl2}
\end{eqnarray}
Equation (\ref{eqn-eigen-nl2}) can be solved for $y$ explicitly. Substitution of the result
into equation (\ref{eqn-eigen-nl1}) then leads to a single equation for $x$, which can be solved
numerically. In the limit of large $\delta$ and $\epsilon$, the eigenvalues are
easily found by nonlinear optimization with the linear eigenvalues as initial
guesses.

Similar to the case of two levels \cite{Smer97}, the nonlinear three-level
model can also be described as a classical Hamiltonian system.
Introducing the variables $p_1 = |a|^2$, $p_3 = |c|^2$,
$q_1 = \arg(b)- \arg(a)$ and $q_3 = \arg(b)- \arg(c)$,
the dynamics is given by the conjugate equations
\be
  \dot p_j = - \frac{\partial \cal H}{\partial q_j} \quad \mbox{and} \quad
  \dot q_j = \frac{\partial \cal H}{\partial p_j}
\ee
with the classical Hamiltonian function
\begin{eqnarray}
  {\cal H} && \!\!\!\! = \epsilon p_1 + \delta p_3
   + \dfrac{g}{2} \left (p_1^2 + p_3^2 + (1-p_1-p_3)^2 \right)
   \label{eqn-classical-hamiltonian} \\
   && +2 \sqrt{1-p_1-p_3}
   \left(v\sqrt{p_1} \cos(q_1) + w \sqrt{p_3} \cos(q_3) \right) \nonumber
\end{eqnarray}
and the normalization condition $|b|^2 = 1 - |a|^2 - |c|^2$.
The eigenstates (\ref{eqn-def-eigenstates}), resp. the stationary
states of the system, are given by $\dot q_j = \dot p_j = 0$.
Hence they correspond to the critical points of this classical Hamiltonian,
which one finds to be given by the real roots of two polynomials in $p_1$
and $p_3$, one of them of 8th order in $p_1$ and of 7th in $p_3$ the other
one the opposite way round.
At these critical points, the classical Hamiltonian and the chemical
potential $\mu$ are simply related by
\be
{\cal H} =\mu - \dfrac{g}{2} \left(p_1^2+p_3^2 + (1-p_1-p_3)^2 \right).
\ee

New eigenstates emerge when the nonlinearity $|g|$ is increased,
starting from three eigenstates in the linear case $g=0$.
From the point of view of standard quantum mechanics, the existence
of additional eigenstates is a strange issue.
Simply considering them as stationary states is a more adequate point
of view.
One can intuitively understand that a strong interaction compared to
the on-site-energies not only modifies the stationary states of the
linear case but can stabilize other states as well and thus create
additional eigenstates. For a deeper insight in the features of these
states see, e.g., \cite{Dago02}, where the nonlinear eigenstates
in a symmetric double-well trap are studied in detail.

In terms of the corresponding Hamiltonian system no conflict arises at
all. The (dis)appearance of critical points
due to a variation of the system parameters is a well-known issue in
the theory of classical dynamical systems. In fact this is a
major ingredient of catastrophe theory \cite{Post78}.
Another important results from the theory of classical dynamical
systems is that the number of extrema minus the number of saddle points
of ${\cal H}$ is constant (see, e.g. \cite{Arno80})
Thus two fixed points, one elliptic and one hyperbolic, always
emerge together.

Furthermore this classical description is especially suited for the
discussion of adiabatic processes \cite{Liu03}.
Note that the dynamics of the discrete nonlinear Schr\"odinger equation
can become classically chaotic \cite{Thom03}.

\section{Landau-Zener tunneling}
\label{sec-zener}

\begin{figure}[t]
\centering
\includegraphics[width=8cm,  angle=0]{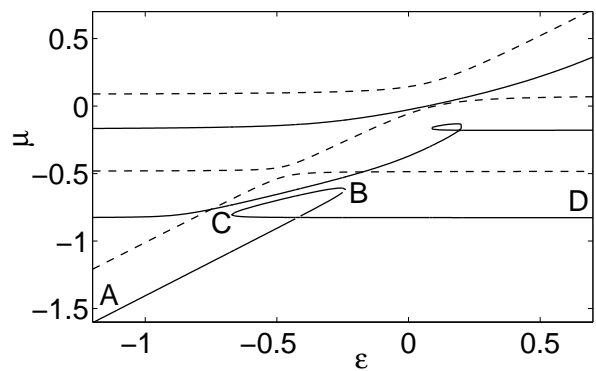}
\caption{\label{fig-loops1}
Nonlinear eigenvalues as defined by equation (\ref{eqn-def-eigenstates})
in dependence of $\epsilon$ for $\delta = -0.4$, $v= 0.1$, $w=0.2$ and
$g = -0.4$ (solid line) resp. $g = 0$ (dashed line).}
\end{figure}

In this section we study the evolution of the system under variation of the
parameters in the manner of the Landau-Zener model, considering $v$
and $w$ independent of time.
In particular we will focus on the so called equal-slope case \cite{Brun93},
for which $\delta$ is constant in time as well and $\epsilon=\alpha t$.
In the linear case $g=0$ the instantaneous eigenvalues in dependence
of $\epsilon =\alpha t$ (the so called adiabatic levels)
form two consecutive avoided crossings which appear
at $\epsilon=\lambda_{\pm}$ with gaps of size
$v_{1,2}=-{\lambda_{1,2}v}/{n_{2,1}}$
with $\lambda_{1,2}= \delta/2 \pm \sqrt{\delta^2/4+w^2}$
and $n_{1,2}=\sqrt{\lambda_{2,1}^2+w^2}$.
An example is shown in figure \ref{fig-loops1} (dashed lines).
If the system is prepared in a state on the lowest branch for
$t \rightarrow -\infty$ and the parameters are varied infinitely slow
(i.e. $\alpha \rightarrow 0$) with $g=0$, the adiabatic theorem states
that the quantum state will follow the adiabatic eigenstate up to a global
phase \cite{Mess99}. However, for a finite value of $\alpha$, parts
of the population will tunnel to the other adiabatic levels, mainly at
the avoided crossings. In the case of the linear two-level system the
transition probability $P$ between the two adiabatic levels is given by the
celebrated Landau-Zener formula \cite{Zene34}
\be
  P_{LZ}=\exp{\left(- {2 \pi v^2}/{\alpha}\right)} \, .
  \label{eqn-LZ-rate}
\ee
For the equal-slope case in linear three-level systems, one can show that
the transition probability to the highest adiabatic eigenstate
(or equivalently the survival probability in the diabatic eigenstate
populated initially) is also given by the Landau-Zener-formula
(\ref{eqn-LZ-rate}) and consequently independent of $w$ and $\delta$
(see \cite{Demk66,Demk68,Brun93} and references therein).

\begin{figure}[thb]
\centering
\includegraphics[width=8cm,  angle=0]{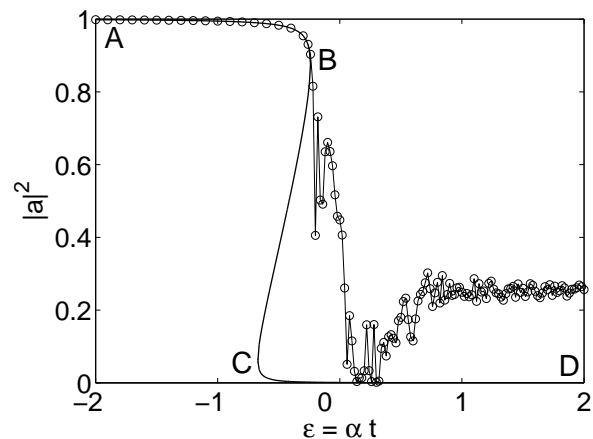}
\caption{\label{fig-zener-dyn}
Time evolution of the population in the first level ($|a(t)|^2$, open circles
$\circ-$) in comparison with the first component of the corresponding
adiabatic eigenstates ($|a_{\rm ad}|^2$, solid line) for
$\epsilon = \alpha t$ with $\alpha = 0.001$ and $\delta = -0.4$, $v= 0.1$, $w=0.2$
and $g=-0.4$.}
\end{figure}

In the following we study the behavior of the system in the equivalent scenario with
${g}\neq{0}$. For that purpose we first calculate the nonlinear eigenvalues,
which are defined by equation (\ref{eqn-def-eigenstates}).
As in the two-level case \cite{Wu00}, loops emerge with increasing nonlinearity near the points
of the avoided crossings at critical values of $|g|=g_{c1}$ resp. $|g| = g_{c2}$,
which are governed by the width of the gap at the avoided crossings in the linear
case.
An estimate of the critical values $g_{c1}$ and $g_{c2}$ is obtained by
considering the two avoided crossings as independent and hence neglecting
the influence of the remaining third level.
For a two-level system, the critical nonlinearity can be calculated exactly
\cite{Wu00}, which yields $g_{c1}=2v_1$ and $g_{c2}=2v_2$. Our numerical studies
show that this is an acceptable approximation and
at least a lower bound to the real value. An example of the nonlinear
eigenvalues in dependence of $\epsilon$ is shown in figure \ref{fig-loops1}
for $\delta = -0.4$, $v = 0.1$ and $w = 0.2$.
We choose $g = -0.4 < 0$ such that the loops emerge on top of the two lower
adiabatic levels for better comparison with \cite{Wu00}. However, similar results
are found if the signs of both $g$ and $\alpha$ are altered.

The first component of the nonlinear eigenstates associated to the
lowest level in figure \ref{fig-loops1} are illustrated in figure
\ref{fig-zener-dyn} (solid line).
The loop structure of the eigenvalues manifests in an S-shaped structure
of the components of the corresponding eigenstate \cite{Zoba00}.
For a better comparison between the figures \ref{fig-loops1} and
\ref{fig-zener-dyn} we included the labels A-D.
The nonlinear eigenstates emerge/vanish in a bifurcation at edges of
the loops resp. the S-structure (marked by B and C in the figures).

As in the two-level system, the appearance of the loops leads to the breakdown
of adiabatic evolution. To illustrate this issue we calculate the dynamics for
the same parameters as above with a slowly varying $\epsilon = \alpha t$ with
$\alpha = 0.001$ and compare them to the relevant instantaneous eigenstates.
The system is initially ($t \rightarrow - \infty$) prepared in an eigenstate
corresponding to the lowest level in figure \ref{fig-loops1} (point A).
The resulting dynamics of the population in the first level $|a(t)|^2$ is shown
in figure \ref{fig-zener-dyn} (open circles). The solid line shows the population
in the first well $|a_{\rm ad}|^2$ for the corresponding instantaneous eigenstates.
One clearly observes the breakdown of the adiabatic evolution when the adiabatic
eigenstate vanishes in a bifurcation at the edge of the S-structure at the point
B around $\alpha t \approx -0.25$.

\begin{figure}[t]
\centering
\includegraphics[width=8cm,  angle=0]{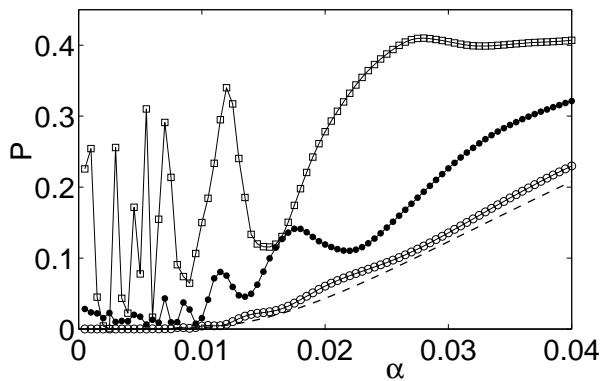}
\caption{\label{fig-zenerp_varg1}
Tunneling probability $P(\alpha)$ in dependence of the velocity
of the parameter variation for $g = -0.03$ (open circles),
$g=-0.16$ (full circles) and $g=-0.4$ (open squares) and
$\delta = -0.4$, $v = 0.1$, $w = 0.2$.
The solid lines are drawn to guide the eye. The Landau-Zener
formula (\ref{eqn-LZ-rate}) for the linear case is shown
for comparison (dashed line).}
\end{figure}

Due to this breakdown of adiabaticity the transition probability $P$
does {\it not} vanish for $\alpha \rightarrow 0$ for $|g| > g_c$. This is
illustrated in fig.~\ref{fig-zenerp_varg1}, where we plotted $P$ in
dependence of $\alpha$ for different values of the nonlinear parameter $g$.
For weak nonlinearities, e.g. $g=-0.03$ in the figure, $P$ is increased
in comparison to the linear case, however it still tends to zero in the
adiabatic limit, i.e.~$P\rightarrow 0$ for $\alpha \rightarrow 0$, as no
loops have occurred yet.
This is no longer true after the appearance of the two loops shown in
fig.~\ref{fig-loops1} such that $P(\alpha \rightarrow 0) > 0$
(cf.~fig.~\ref{fig-zenerp_varg1}).
These features are well-known from the nonlinear two-level model \cite{Wu00}.
However, a novel feature is the appearance of pronounced oscillations of the
transition probability $P(\alpha)$ for small values of $\alpha$ due to the
nonlinear interaction between the different levels.

\begin{figure}[t]
\centering
\includegraphics[width=8cm,  angle=0]{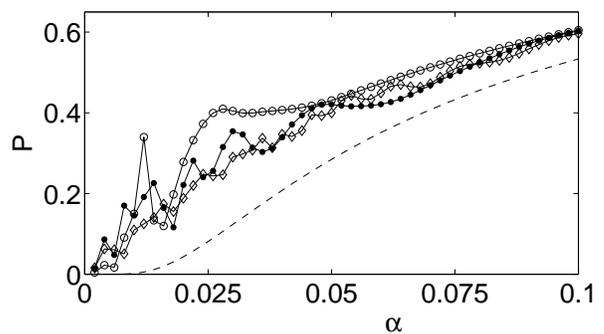}
\caption{\label{fig-zenerp_varw1}
Tunneling probability $P(\alpha)$ in dependence of the velocity
of the parameter variation for $w = 0.2$ (open circles),
$w=0.4$ (full circles) and $w=0.6$ (open squares) and
$\delta = -0.4$, $v=0.1$ and $g=-0.4$.
The solid lines are drawn to guide the eye. The Landau-Zener
formula (\ref{eqn-LZ-rate}) for the linear case is shown
for comparison (dashed line).}
\end{figure}

As mentioned above, the transition probability for the linear case is
independent of $w$ and $\delta$.
This does not hold in the nonlinear case any longer, since the influence
of the nonlinearity, e.g.~the emergence and the structure of the loops,
obviously depends on $v, \, w$ and $\delta$.
This is confirmed by our numerical results
shown in fig.~\ref{fig-zenerp_varw1}, where $P(\alpha)$ is plotted
for different values of $w$.
Again pronounced oscillations of $P(\alpha)$ are found for small
values of $\alpha$.
These oscillations and their dependence on the system's parameters
will be studied in detail in a subsequent paper.

\section{Nonlinear STIRAP}
\label{sec-stirap}

\begin{figure}[t]
\centering
\includegraphics[width=8cm,  angle=0]{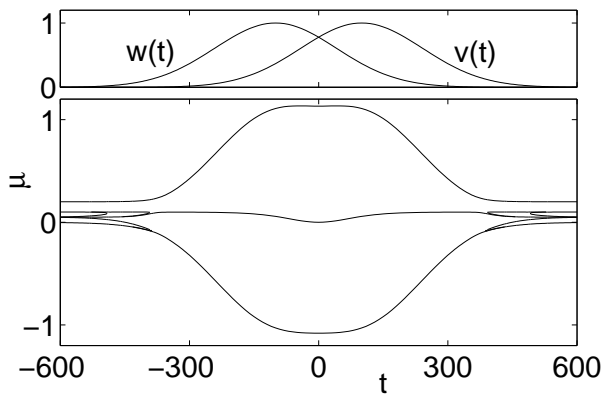}
\caption{\label{fig-stirap-mu1}
Eigenvalues of the nonlinear Hamiltonian (\ref{eqn-nlham-intro}) for
a detuning $\Delta = 0.1$ and a nonlinearity $g=0.2$ (lower panel)
for time-dependent couplings $v(t)$ and $w(t)$ (upper panel).}
\end{figure}

The STIRAP method, primarily proposed and realized for atomic
three-level systems \cite{Berg98b}, allows a robust
coherent population transfer between quantum states.
In the meanwhile STIRAP has been generalized to systems with
multiple levels and the preparation of coherent superposition
states \cite{Vita01}.
In this case the coupling between the atomic bare states
are realized by slightly detuned time-dependent laser fields
with Rabi frequencies $v(t)$ and $w(t)$.
In the rotating wave approximation at the two-photon resonance,
the dynamics of the three-level atom is given by the Hamiltonian
(\ref{eqn-nlham-intro}) with $g = 0$ and a fixed detuning
$\Delta = -\delta = -\epsilon$. Using the STIRAP scheme one can
achieve a complete population transfer from level $\psi_1$ to
level $\psi_3$ by an adiabatic passage via a dark state
$\psi_{\rm ds}$ of the system, which is a superposition
of $\psi_1$ and $\psi_3$ alone.
If the coupling $w(t)$ between the levels $\psi_2$ and $\psi_3$
is turned on {\it before} the coupling $v(t)$ between the levels
$\psi_1$ and $\psi_2$, the system's dark state is rotated from
$\psi_{\rm ds}(t= - \infty) = \psi_1$ to $\psi_{\rm ds}(t = + \infty)) = \psi_3$.
If the parameters $v(t)$ and $w(t)$ are varied sufficiently slowly,
the system can follow the dark state adiabatically  which leads
to a complete population transfer from level $\psi_1$ to $\psi_3$.
This counterintuitive pulse sequence of the coupling
elements $v(t)$ and $w(t)$ is illustrated in fig.~\ref{fig-stirap-mu1}
(upper panel).

\begin{figure}[t]
\centering
\includegraphics[width=8cm,  angle=0]{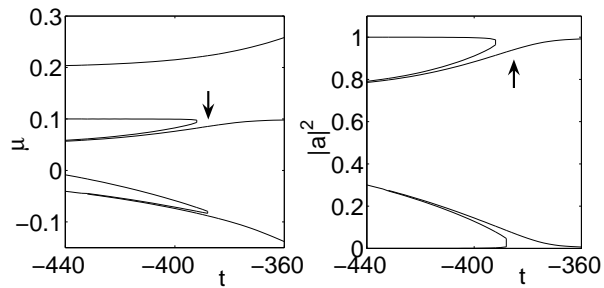}
\caption{\label{fig-stirap-mua_crossing}
Left: Magnification of fig.~\ref{fig-stirap-mu1}.
Right: Squared modulus of the first component ($|a_{\rm ad}|^2$) of the
corresponding eigenstates. The avoided crossing that leads to
a breakdown of adiabaticity is marked by an arrow.}
\end{figure}

\begin{figure}[t]
\centering
\includegraphics[width=8cm,  angle=0]{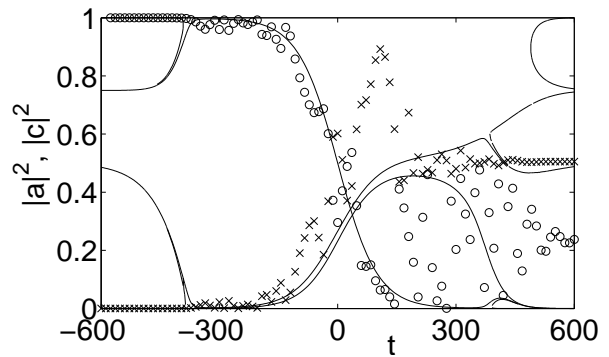}
\caption{\label{fig-stirap-dyn}
Time evolution of the population in the first ($|a(t)|^2$, open circles)
and the third level ($|c(t)|^2$, crosses) in comparison with the population
in the first level for the instantaneous eigenstates ($|a_{\rm ad}|^2$, solid
lines) for $\Delta = 0.1$ and $g=0.2$ and the couplings $v(t),w(t)$ shown
in fig.~\ref{fig-stirap-mu1} (upper panel).}
\end{figure}

However, the situation is more involved in the nonlinear case.
To begin with we consider the case $g,\Delta > 0$.
As noted in the previous section, new nonlinear eigenstates
emerge if the nonlinearity $|g|$ exceeds a critical value $g_c$ depending
on the other sytem parameters. This can give rise to new crossing scenarios
leading to a breakdown of adiabaticity.
To illustrate this issue, the eigenvalues and eigenstates are calculated
for a detuning $\Delta = 0.1$ and $g=0.2$.
The resulting adiabatic eigenvalues as well as the STIRAP pulse sequence
of the couplings $v(t)$ and $w(t)$ are shown in fig.~\ref{fig-stirap-mu1}.
At a first glance this picture looks quite similar to the linear case.
The levels are shifted slightly due to the mean-field energy and
a few additional nonlinear eigenvalues emerge for large $|t|$, because then
the coupling elements $v$ and $w$ are small compared to the nonlinearity.
However, a closer look at the adiabatic eigenvalues and eigenstates
shown in fig.~\ref{fig-stirap-mua_crossing} reveals a fatal nonlinear
avoided crossing scenario around $t \approx -380$, which will be referred to
as the (avoided) ''horn crossing'' in the following because of its shape.
In the linear case the system's dark state is rotated from 
$\psi_{\rm ds}(t= - \infty) = \psi_1$ to 
$\psi_{\rm ds}(t = + \infty)) = \psi_3$, which leads to a coherent
adiabatic population transfer.
In the horn crossing however, this ''dark state'' disappears when it
merges with a nonlinear eigenstate (which will be referred to as 
horn state in the following), such that no adiabatic passage is 
possible any longer. 
To illustrate this breakdown of nonlinear STIRAP, we integrate the
three-level NLSE numerically for the same parameters as in
fig.~\ref{fig-stirap-mu1}. The resulting evolution of the
populations $|a(t)|^2$ and $|c(t)|^2$ is shown in
fig.~\ref{fig-stirap-dyn} in comparison to the population
$|a_{\rm ad}|^2$ of the instantaneous eigenstates.
Due to the crossing, a state initially prepared in level
$\psi_1$, $a(t=-\infty) = 1$, cannot be transfered to level
$\psi_3$ adiabatically any more.
The dynamics follows the instantaneous eigenstate (the ''dark
state'') adiabatically until this state disappears at the horn
crossing. Fast oscillations of the populations $|a(t)|^2$ and
$|c(t)|^2$ are observed afterwards. Finally the system settles
down to a steady state again, but the population has not been
transferred completely.

\begin{figure}[t]
\centering
\includegraphics[width=8cm,  angle=0]{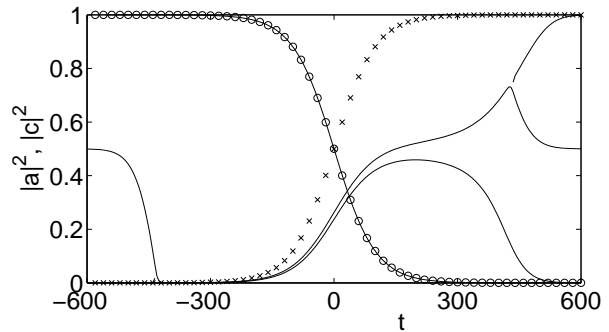}
\caption{\label{fig-stirap-dyn005}
As fig.~\ref{fig-stirap-dyn}, however for $g=+0.05$.}
\end{figure}

For $g,\Delta > 0$, it is found that the nonlinear eigenstate (the horn 
state) that merges with the dark state is a superposition of $\psi_1$ 
and $\psi_2$ alone in the limit $t \rightarrow -\infty$ and 
consequently $v,w \rightarrow 0$. Substitution of $c = 0$ and 
$v = w = 0$ into equation (\ref{eqn-def-eigenstates}) immediately 
leads to
\be
  - \Delta + g|a|^2 = \mu = g|b|^2.
\ee
Solving for $\Delta$ yields a condition for the existence of the horn state,
\be
  \Delta = g(|a|^2-|b|^2) \le g.
\ee
In this case a complete population transfer using the STIRAP scheme will
be prevented by the horn crossing scenario discussed above.
Nonlinear STIRAP is still possible if the nonlinearity is smaller than the
detuning, $g < \Delta$.
This is illustrated in fig.~\ref{fig-stirap-dyn005}, where the evolution of
the populations $|a(t)|^2$ and $|c(t)|^2$ is plotted in comparison to the
population $|a_{\rm ad}|^2$ of the adiabatic eigenstates for $g=0.05$ and
$\Delta = 0.1$. One observes that the dynamics closely follows the adiabatic
eigenstate and that the population is transferred from $\psi_1$ to $\psi_3$
completely.
Note that the probability of Landau-Zener tunneling is also
increased for $g < \Delta$ if the system parameters are varied at a finite
velocity (cf. section \ref{sec-zener} and ref. \cite{Wu00}).

\begin{figure}[t]
\centering
\includegraphics[width=8cm,  angle=0]{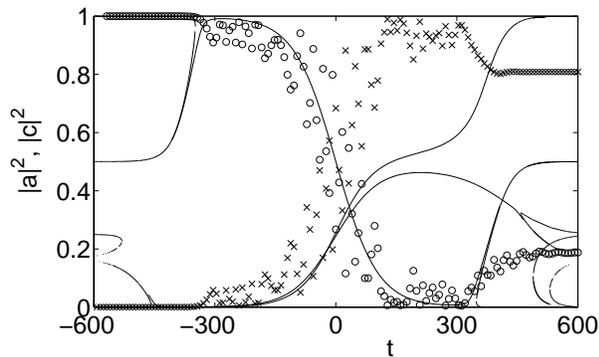}
\caption{\label{fig-stirap-dyn-2}
As fig.~\ref{fig-stirap-dyn}, however for $g=-0.2$.}
\end{figure}

\begin{figure}[t]
\centering
\includegraphics[width=8cm,  angle=0]{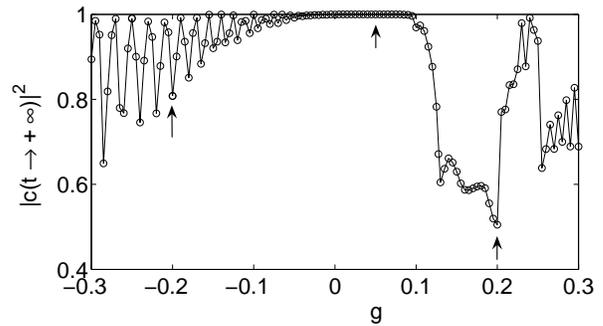}
\caption{\label{fig-stirap-eff}
Numerically calculated transfer efficiency of nonlinear STIRAP
in dependence of the nonlinearity $g$ for $\Delta = 0.1$.
The arrows indicate the values of $g$ for which the dynamics is shown
in fig.~\ref{fig-stirap-dyn}, \ref{fig-stirap-dyn005}
and \ref{fig-stirap-dyn-2}, respectively.}
\end{figure}

An analogous situation is found for $g,\Delta < 0$.
However, a different crossing scenario arises if the signs of $\Delta$
and $g$ are opposite. Again the ''dark state'' disappears when it merges
with a nonlinear eigenstate (horn state). In this case it is found that the
horn state is now a superposition of $\psi_1$ and $\psi_3$ alone in the
limit $t \rightarrow - \infty$ as the dark state and that it exists for
all values of $g$.
Rigorously speaking, this leads to a breakdown of STIRAP even for very
small nonlinearities. However, this crossing turns out to be not as
fatal as the one discussed above; the transfer is still close to unity
for weak nonlinearities (cf. fig.~\ref{fig-stirap-eff}). 
An example of the dynamics is  shown in fig.~\ref{fig-stirap-dyn-2} 
for $\Delta = 0.1$ and $g=-0.2$.

In conclusion we find the following conditions for the feasibility
of a complete adiabatic population transfer using the STIRAP scheme:
\be
  g \Delta \ge 0 \quad \mbox{and} \quad |g| < g_c = |\Delta|
\ee
The dependence of the transfer efficiency $|c(t \rightarrow + \infty)|^2$
on the nonlinearity is shown in fig.~\ref{fig-stirap-eff} for
$\Delta = 0.1$ and the same couplings $v(t)$ and $w(t)$ used in
fig.~\ref{fig-stirap-mu1} and \ref{fig-stirap-dyn}.
Note that exactly the same dependence is found if the signs of
both $\Delta$ and $g$ are altered.
The transfer efficiency is slightly reduced for all values $g < 0$
and shows an oscillatory behavior.
For $g > 0$, one clearly observes an abrupt breakdown of the transfer
efficiency above the critical nonlinearity, $g \ge g_c = |\Delta|$.\\

\section{Conclusion}
\label{sec-conclusion}

In conclusion, the eigenstates and the dynamics generated by
the nonlinear Hamiltonian (\ref{eqn-nlham-intro}) are analyzed 
for two important cases:
the equal-slope Landau-Zener model and the STIRAP scheme.
The emergence of new nonlinear eigenstates and novel crossing scenarios
leads to a breakdown of adiabatic evolution if the nonlinearity $|g|$ exceeds
a critical value.
Consequently, STIRAP fails if the nonlinearity exceeds a critical value
given by the detuning or if the nonlinear parameter and detuning have
different signs. A novel feature of nonlinear Zener tunneling
compared to the two-level system is the
oscillatory behavior of the transition probability $P(\alpha)$.
Open problems include a detailed analysis
of the oscillations of $P(\alpha)$, the Landau-Zener scenario for
non-equal slope and the effects of a parameter variation at finite
velocity on the nonlinear STIRAP scheme.\\

\begin{acknowledgments}
Support from the Deutsche Forschungsgemeinschaft
via the Graduiertenkolleg ''Nichtlineare Optik und Ultrakurzzeitphysik''
is gratefully acknowledged. We thank B.~W.~Shore, U.~Schneider and
K. Bergmann for stimulating discussions.
\end{acknowledgments}


\end{document}